\documentclass[10pt, aps, prd, superscriptaddress, nofootinbib, showpacs, twocolumn, longbibliography]{revtex4-2}
\usepackage[english]{babel}
\usepackage{amsmath, amssymb, amsfonts, amsthm, mathrsfs}
\usepackage[all]{xy}
\usepackage{bm}
\usepackage{stmaryrd}
\usepackage{graphicx}
\usepackage{hyperref}
\hypersetup{
breaklinks=true,   
}

\usepackage{color}

\usepackage{pgfplots, grffile, tikz}
\pgfplotsset{compat=newest}
\usetikzlibrary{plotmarks, arrows.meta}
\usepgfplotslibrary{patchplots}

\numberwithin{equation}{section}

\newcommand{\lp}{\left(}
\newcommand{\rp}{\right)}
\newcommand{\lb}{\left[}
\newcommand{\rb}{\right]}
\newcommand{\lc}{\left\{}
\newcommand{\rc}{\right\}}

\DeclareMathOperator{\Det}{\mathrm{Det}}
\DeclareMathOperator{\Tr}{\mathrm{Tr}}
\DeclareMathOperator{\tr}{\mathrm{tr}}

\newcommand{\pp}{\partial}

\newcommand{\h}{\hat}
\renewcommand{\t}{\widetilde}

\renewcommand{\cal}{\mathcal}

\renewcommand{\a}{\alpha}
\renewcommand{\b}{\beta}
\newcommand{\g}{\gamma}
\renewcommand{\d}{\delta}

\renewcommand{\l}{\lambda}
\newcommand{\m}{\mu}
\newcommand{\n}{\nu}

\newcommand{\vk}{\varkappa}

\renewcommand{\O}{\varOmega}
\newcommand{\G}{\Gamma}

\newcommand{\nb}{\nabla}

\newcommand{\calR}{\mathcal{R}}
\newcommand{\calO}{\mathcal{O}}

\newcommand{\co}[1]{C_{#1}}

\newcommand{\ph}{\phantom}

\begin{document}

\title{Notes on peculiarities of Schwinger--DeWitt technique: one-loop double poles, total-derivative terms and determinant anomalies}

\author{A.\,O. Barvinsky}
\email{barvin@td.lpi.ru}
\affiliation{Theory Department, Lebedev Physics Institute, Leninsky Prospect 53, Moscow 119991, Russia}
\author{A.\,E. Kalugin}
\email{kalugin.ae@phystech.edu}
\affiliation{Theory Department, Lebedev Physics Institute, Leninsky Prospect 53, Moscow 119991, Russia}

\begin{abstract}
We discuss peculiarities of the Schwinger--DeWitt technique for quantum effective action, associated with the origin of dimensionally regularized double-pole divergences of the one-loop functional determinant for massive Proca model in a curved spacetime. These divergences have the form of the total-derivative term generated by integration by parts in the functional trace of the heat kernel for the Proca vector field operator. Because of the nonminimal structure of second-order derivatives in this operator, its vector field heat kernel has a nontrivial form, involving the convolution of the scalar d'Alembertian Green's function with its heat kernel. Moreover, its asymptotic expansion is very different from the universal predictions of Gilkey-Seeley heat kernel theory because the Proca operator violates one of the basic assumptions of this theory --- the nondegeneracy of the principal symbol of an elliptic operator. This modification of the asymptotic expansion explains the origin of double-pole total-derivative terms. Another hypostasis of such terms is in the problem of multiplicative determinant anomalies---lack of factorization of the functional determinant of a product of differential operators into the product of their individual determinants. We demonstrate that this anomaly should have the form of total-derivative terms and check this statement by calculating divergent parts of functional determinants for products of minimal and nonminimal second-order differential operators in curved spacetime.
\end{abstract}

\maketitle

\section{Introduction}
\label{sec:intro}
Functional determinants play a significant role in many areas of theoretical physics, but they are of especially high importance in the background field formalism of quantum gauge theories and quantum gravity~\cite{schwinger_orig,dewitt_book} wherein they appear in many applications. In particular, they serve as one-loop quantum effective action $\varGamma_{\rm one-loop}$ on generic gravity and matter backgrounds and turn out to be critically important for the analysis of renormalizability of gravitational and matter field models~\cite{Hooft-Veltman}. The Schwinger-DeWitt technique of their calculation is based on the heat kernel method whose rigorous mathematical foundations rely on Gilkey-Seeley theory~\cite{seeley_orig,gilkey_orig} for differential operators $F$---inverse propagators of physical fields on smooth background manifolds. The essence of this technique consists in the asymptotic expansion of the heat kernel of such operators $F$,
\begin{equation}
K_F(\tau\,|\,x,y)\equiv e^{\tau F}\delta(x,y)
\end{equation}
in the limit of small proper time parameter $\tau$. For the Hermitian differential operator $F$ of order $2M$ acting in $d$-dimensional space $\cal{M}$ and having an {\em invertible} positive definite principal symbol (precise definitions will be given later) the asymptotic expansion for the functional trace of their heat kernel at $s\to 0$ has the form
\begin{equation}\label{Gilkey-S}
{\rm Tr}\, e^{\tau F}\equiv\int\limits_{\cal M} d^dx\,K_F(\tau\,|\,x,y)\,\big|_{\,y=x}
=\frac1{\tau^{d/2M}}\sum\limits_{n=0}^\infty A_n\,\tau^{n/2M}.
\end{equation}
The coefficients $A_n$ of this expansion are given by the sum of the volume integrals over the spacetime $\cal M$ and surface integrals over the boundary $\partial\cal M$,
\begin{equation}
A_n=\int\limits_{\cal M} d^dx\,E_n(x)+\int\limits_{\partial\cal M} d^{d-1}x\,B_n(x).  \label{A_n}
\end{equation}

Spacetime volume densities
\begin{equation}
E_{2m}(x)=\frac{a_m(x)}{(4\pi)^{d/2k}}  \label{E2m}
\end{equation}
are non-zero only for even $n=2m$, $m=0,1,...$; $E_{2m+1}(x)=0$, and they are built in terms of the coefficients of the differential operator, while the surface densities $B_n(x)$ are, in general, non-zero for both odd and even $n$, and critically depend on the boundary conditions for the operator $F$ on $\partial\cal M$. Apart from the coefficients of $F$ they essentially depend on such characteristics of the boundary $\partial\cal M$ as its extrinsic curvature.

The Gilkey-Seeley method was actually preceded in mathematics and physics by the works of Hadamard, Minakshisundaram~\cite{Hadamard, Minakshisundaram} and the Schwinger-DeWitt technique~\cite{dewitt_book} dealing with the spacetime volume part of (\ref{A_n}), which is why the term HaMiDeW coefficients was coined for $a_m(x)$ in (\ref{E2m}). This technique turned out to be highly efficient in field theory as a calculational tool for quantum effects in external matter and gravitational fields, and the goal of the present work will be to clarify several important issues within this method, which we formulate below.

The first important issue is associated with the fact that the calculation of surface terms, unlike their volume counterparts, is much more involved and not so universal. This is because it is strongly biased by specific boundary conditions for the operator $F$. At the same time the volume part of $A_n$ contains total derivative terms under the integral sign. Upon integration these terms reduce to the surface contributions anyway. Those surface contributions, however, are independent of both boundary conditions and the properties of the boundary. Thus, they are rather universal and play an important role in such objects as conformal anomalies.\footnote{In fact, these total derivative terms can be extracted from the integral of $E_{2(m+M)}(F+\varphi\,|x)$ for the $2M$-th order operator $F(\nabla)+\varphi$ modified by the addition of the function $\varphi=\varphi(x)$ with a compact support, $F(\nabla)\to F(\nabla)+\varphi$. This can be done via a simple variational equation,
\begin{equation}
E_{2m}(F|x)=\frac\delta{\delta\varphi(x)}\int d^dy\,E_{2(m+M)}(F+\varphi\,|y)\,\Big\vert_{\varphi=0},
\end{equation}
which is equivalent to smearing the integrand of (\ref{Gilkey-S}) with the function $\varphi(x)$.} For example, in four dimensional models a total derivative contribution $\Box R$ to $E_4(x)$ ($R$ being a Ricci scalar and $\Box=g^{\mu\nu}\nabla_\mu\nabla_\nu$ being the covariant d'Alembertian built of the metric $g_{\mu\nu}$ and covariant derivatives $\nabla_\mu$ with the metric preserving fibre bundle connection) is a part of the trace anomaly for quantum stress tensor of the classically conformally-invariant field. Thus, we will not be dealing with the $B_n$ part of (\ref{A_n}), which is beyond our control unless the boundary value problem for $F$ is precisely specified, but instead will focus on $\Box R$-type contributions to $E_4(x)$ in $d=4$ case.

For the so-called minimal second order operators whose derivatives form the d'Alembertian, $F(\nabla)=\Box+P$, the contribution of such terms is well-known and follows from the Schwinger-DeWitt expansion for the heat kernel of such operators~\cite{dewitt_book,barvin1985}
\begin{multline} \label{HeatKernel0}
K_F(\tau|x,y) = \frac{\Delta^{1/2}(x,y)}{(4\pi\tau)^{d/2}}\,g^{1/2}(y)
\sum\limits_{m=0}^\infty \tau^{m}\\
\times \exp\left(-\frac{\sigma(x,y)}{2\tau}\right)\, a_m(x,y).
\end{multline}
Here  $\sigma(x,x')$ is the Synge world function---one half of the square of the geodetic distance between the points $x$ and $y$, $\Delta(x,y)=\det(-\partial^2\sigma)/\partial x^\mu\partial y^\nu)/g^{1/2}(x)\,g^{1/2}(y)$  is the dedensitized Pauli-Van Vleck-Morette determinant and $a_m(x,y)$ are the two-point HaMiDeW or Schwinger-DeWitt coefficients whose coincidence limits $a_m(x)=a_m(x,x)$ are featuring in Eq.(\ref{E2m}). These coefficients satisfy recurrent differential equations which can be successively solved for $\hat a_m(x,y)$ in the form of covariant Taylor series in powers of geodetic separation between $x$ and $y$. The coefficients of this Taylor expansion are local functions of spacetime metric, its curvature and background fields, and thus provide all the goals of perturbative UV renormalization of local field models and their effective field theory expansion.

For nonminimal operators with a complicated structure of highest order derivatives the calculational algorithm is less universal and usually circumvents the use of the heat kernel by replacing it with the method of universal functional traces~\cite{barvin1985}. The simplest example of the non-minimal operator is the vector field operator of Proca theory which is very interesting in view of the observation made in~\cite{barvin1985}. There the calculation of one-loop effective action for the Proca field has led within dimensional regularization to double-pole UV divergences which also turned out to be given by the total-derivative $\Box R$-terms. But this is impossible for models subject to Gilkey-Seeley method, because the divergent part of the one-loop effective action in spacetime of even dimension $D$ according to (\ref{Gilkey-S}) is always given by the single pole term in the dimension $d\to D-0$ extended to the complex plane,
\begin{equation}
\frac12{\rm Tr}\,\ln F\,\Big|^{\rm div}=-\frac12\int\limits_0^\infty\frac{d\tau}\tau\,{\rm Tr}\, e^{\tau F}\,\Big|^{\rm div}
=-\frac{M\,A_{D/2}}{D-d},  \label{div}
\end{equation}
where $M$ determines the highest power $2M$ of derivatives in the operator $F$---the inverse propagator of the theory.

Our goal in Sect.\ref{sec:2pole} will be to pinpoint the origin of this contradiction and reveal its mechanism. As will be shown, this property is the artefact of the violation of non-degeneracy of the operator, one of the assumptions of the Gilkey--Seeley theory. Proca operator is invertible as a whole and has a well defined Green function, but its invertibility is achieved in virtue of the non-zero mass term, something which goes beyond the assumptions of Gilkey-Seeley theory. As a by-product of this analysis we will explicitly construct the heat kernel for the nonminimal Proca field operator and confirm our explanation of the double-pole UV divergence by the resummation method explicitly disentangling the total derivative double pole term.

In Sect.\ref{sec:det_anomalies} we will consider the so-called problem of the functional determinant anomaly. As is well known, functional determinants do not fully replicate the properties of finite dimensional matrices. Due to infinite dimensionality of the operator functional spaces, the functional determinant of the product of differential operators is not guaranteed to be equal to the product of their individual determinants~\cite{kontsevich1994, gilkey_anom, wodz_book}.
Although it has been argued that such anomalies are unphysical and are merely an artefact of $\zeta$-regularization~\cite{evans_unphys}, there does not seem to be a consensus on this matter~\cite{bytsenko_book, elizade_co1, elizade_co2}. For some differential operators ${\mathcal{O}}_1$ and ${\mathcal{O}}_2$ one can define, following~\cite{bytsenko_book}, the quantity
\begin{equation}
\mathcal{A}({\mathcal{O}}_1,{\mathcal{O}}_2)\!=
\ln\Det{(\mathcal{O}_{1}\mathcal{O}_{2})}-\ln\Det{\mathcal{O}}_1\!-\ln\Det{\mathcal{O}}_2,
\label{anomaly_def}
\end{equation}
to which we will further refer as the determinant anomaly for operators ${\mathcal{O}}_1$ and ${\mathcal{O}}_2$. Despite the existence of a rigorous mathematical theory~\cite{wodz_book, wodz_comment} regarding~(\ref{anomaly_def}) and even an explicit formula for this quantity (see, e.g.~\cite{kontsevich1994}), the determinant anomaly remains largely unstudied for physically relevant operators.

Mainly this is due to the fact that the anomaly algorithm relies on the computation of the so-called Wodzicki residue~\cite{kalau_waltze, ackermann} of certain pseudo-differential operators, which is very complicated even for simplest operators on curved manifolds and involves their non-commutative symbol calculus~\cite{kalau_waltze}, or, alternatively, a rather indirect asymptotic expansion of the trace for a product of the aforementioned operator and the heat kernel of an arbitrary elliptic one~\cite{le_wres}.

Thus we will not appeal to the Wodzicki residue theory, but give simple arguments that the violation of this factorization property should occur in the form of total derivative terms and will check this by considering several examples based on fourth order differential operator decomposable into the product of two minimal or nonminimal second order ones. To the best of our knowledge, this fact has not been previously reported in the literature{, although some examples of this property of the anomaly had been previously observed for fermion operators, e.g. in~\cite{shapiro_torsion, shapiro_fermion}}. Such a verification becomes possible for the divergent part of the functional determinants because now we have recently derived algorithms for the Schwinger-DeWitt coefficients of a generic fourth-order differential operator including third order derivatives of the most general form~\cite{4thorder_22,BKW}.

\section{Nonminimal vector field Proca operator}
\label{sec:2pole}
In this section we discuss a seeming contradiction between the Gilkey--Seeley method and the calculation of one-loop effective action in the model of the Proca massive vector field in curved spacetime. The massive Proca vector field operator reads
\begin{equation}
    F\equiv F^{\mu}_{\nu}(m^2|\,\nabla)=\Box\delta^{\mu}_{\nu}-\nabla^{\mu}\nabla_{\nu}-R^{\mu}_{\ph{\mu}\nu}-m^2\delta^{\mu}_{\nu}.
    \label{proca_op_def}
\end{equation}
It is nonminimal, but its Green's function can be expressed in terms of the Green's function of the massive version of the minimal operator $H(\nabla)$,
\begin{equation}\label{H}
H(\nabla)-m^2\equiv H^{\,\m}_{\,\n}(\nabla)-m^2\delta^\mu_\nu=\Box\d^{\m}_{\n}-R^{\m}_{\n}-m^2\delta^\mu_\nu,
\end{equation}
according to a simple relation~\cite{barvin1985}\footnote{One can consider instead of (\ref{proca_op_def}) a more complicated operator with a generic potential term different from $-R^{\m}_{\n}$, for which the Green's function can be built by perturbation theory of the generalized Schwinger-DeWitt technique~\cite{barvin1985}, but for simplicity we will work with the operator (\ref{proca_op_def}) which admits the use of auxiliary Ward identities (see below).}
\begin{equation}\label{GFuntion}
\frac{\delta^{\,\m}_{\,\n}}{F(m^2|\,\nabla)}=\Big(\delta^{\mu}_{\a}-\frac{1}{m^2}\nabla^{\mu}\nabla_{\a}\Big)\,
\frac{\delta^{\,\a}_{\,\n}}{H(\nabla)-m^2},
\end{equation}
where the Kronecker symbol in the numerator of the Green's function indicates its tensor structure as acting on a vector field.\footnote{For brevity, in what follows we will have to use this abbreviation, that was numerously used in previous publications \cite{barvin1985,4thorder_22,Shapiro}, for inversion of differential operators acting on various vector and tensor fields, like $\big[F^{-1}(\nabla)\big]^\mu_\nu=\delta^\mu_\nu/F(\nabla)$, $\big[\Box^{-1}\big]^\mu_\nu=\delta^\mu_\nu/\Box$, etc.}

Modulo usually disregarded volume divergences, $\Tr\ln F(m^2|\nabla)\propto\delta(0)$, $m^2\to\infty$, the one-loop effective action for this operator can be represented as a mass integral of the trace of a massive Green's function
\begin{equation}
\frac{1}{2}\Tr\ln F^{\m}_{\n}(m^2|\nabla)=\frac{1}{2}\Tr\int\limits_{m^2}^{\infty}d\m^2\,\frac{\delta^{\,\m}_{\,\n}}{F(\mu^2|\nabla)},
\label{proca_pt_mu_int}
\end{equation}
\\
Its divergences in dimensional regularization, quite amazingly, contain the double pole term~\cite{barvin1985}
\begin{widetext}
\begin{equation}
\begin{split}
&\frac{1}{2}\Tr\ln F^\mu_\nu(m^2|\nabla)\,\Big\vert^{\mathrm{div}}=\frac{1}{\omega-2}\int \frac{d^4x \,g^{1/2}}{32\pi^2}\bigg\{-\frac{11}{180}R^2_{\a\b\m\n}+\frac{43}{90}R^2_{\a\b}-\frac{1}{9}R^2+\frac{m^2}{2}R+\frac{3}{2}m^4\\
&-\frac{1}{12}(\gamma_E+\ln m^2)\Box R-\frac{1}{30}\Box R\bigg\}-\frac{1}{(\omega-2)^2}\int \frac{d^4x \,g^{1/2} }{32\pi^2}\frac{1}{12}\Box R,\qquad\omega\equiv\frac{d}2\rightarrow2-0.
\end{split}
\label{proca_pt_ans}
\end{equation}
\end{widetext}
This contradicts the Seeley -- Gilkey theory dictating, according to (\ref{div}), that only simple pole-type divergences may be present in this expansion. Moreover, another method of~\cite{barvin1985}, namely the one employing the Ward identities for this operator, did not produce this double-pole term. Here we reproduce this term from the heat kernel obtained by the Green function Mellin transform and show how it also arises in the method of Ward identities on a proper account of surface terms originating from integration by parts.

The above contradiction could have been expected because of a possible breakdown of Gilkey-Seeley expansion due to the fact that the principal symbol of the Proca operator (\ref{proca_op_def}) is degenerate contrary to the assumptions of Gilkey-Seeley theorems~\cite{seeley_orig,gilkey_orig}. Indeed, the matrix of the principal symbol---the term with second order derivatives $\nabla_\mu$ replaced by some vector $ip_\mu$ (in fact the analogue of the Fourier momentum), $p^2\delta^{\mu}_{\nu}-p^{\mu}p_{\nu}$, is not invertible and cannot be used for the construction of the perturbation theory for the Green's function of $F(m^2|\nabla)$. The Green's function constructed above tells us that this perturbation theory should be performed by including the mass term into the extended principal symbol of the operator, $p^2\delta^{\mu}_{\nu}-p^{\mu}p_{\nu}\to D^{\mu}_{\nu}(p)=p^2\delta^{\mu}_{\nu}-p^{\mu}p_{\nu}+m^2\delta^{\mu}_{\nu}$, its inversion ${D^{-1}}^\mu_\nu(p)=(\delta^{\mu}_{\n}+\frac{1}{m^2}p^{\mu}p_{\n})/(p^2+m^2)$ leading to the momentum space analogue of the matrix factor $\delta^{\mu}_{\a}-\frac{1}{m^2}\nabla^{\mu}\nabla_{\a}$ in (\ref{GFuntion}). As a result of this modification, the structure of the heat kernel and its trace expansion becomes different from (\ref{HeatKernel0}) and (\ref{Gilkey-S}), which explains seemingly contradicting results. This will become clear below when we construct the heat kernel for this operator.

\subsection{Heat kernel via Mellin transform and Proca operator Ward identities}
\label{sec:hk_proca}
We will construct the heat kernel of the Proca operator by the Mellin transform of its Green's function. Notice, that for the Proca operator~(\ref{proca_op_def}) the procedure of shifting the mass term into the complex plane yields $F(m^2+z|\nabla)=F(m^2|\nabla)-z$. Here we use this relation to construct the Proca heat kernel $K_{F}(\tau)$. The key fact is that the operator and its heat kernel are related by means of the Mellin (inverse Laplace) integral transform~\cite{hk_manual}. Assuming the existence of a complex plane contour $\cal C$ encircling the spectrum of $F$, we may write

\begin{equation}
\begin{split}
&K^\mu_{\nu, F}(\tau)
=-\frac{1}{2\pi i}\int\limits_{\cal C}dz\,e^{z\tau}\frac{\delta^\mu_\nu}{F(m^2|\nabla)-z}\\
&\qquad\quad\;\;=-\frac{1}{2\pi i}\int\limits_{\cal C}dz\,e^{z\tau}\frac{\delta^\mu_\nu}{F(m^2+z|\nabla)}\\
&=-\frac{1}{2\pi i}\int\limits_{\cal C}dz\,e^{z\tau}\Big(\d^{\m}_{\l}-\frac{1}{m^2+z}\nb^{\m}\nb_{\l}\Big)\frac{\delta^\l_{\,\n }}{H(\nabla)-m^2-z},\\
\end{split}
\label{mellin_tr}
\end{equation}
where we have introduced a minimal vector field operator (\ref{H})

Taking the residue in~(\ref{mellin_tr}), we obtain
\begin{equation}
\begin{split}
K^\mu_{\nu, F}(\tau)&= K^\mu_{\nu, H}(\tau)e^{-m^2\tau}
\\&+\nb^{\m}\nb_{\l}\Big(\,\frac{\delta^\l_\n}{H(\nabla)}-e^{\tau H(\nabla)}\frac{\delta^\l_\n}{H(\nabla)}\,\Big)\, e^{-m^2\tau}.
\end{split}
\label{HKforF}
\end{equation}

To simplify the last two terms let us use what was called in~\cite{barvin1985} ``the Ward identity'' for the vector field operator
\begin{equation}
\begin{split}
\nb_{\a}(\,\Box\d^{\a}_{\b}-R^{\a}_{\b}\,)=\Box_{\rm s}\nb_{\b},
\label{ward}
\end{split}
\end{equation}
where $\Box_{\rm s}$ denotes the covariant d'Alembertian acting on a scalar field (the identity above is certainly understood as applied to the vector field $\varphi^\nu$). This obviously implies that any power of $H$ satisfies a similar property $\nb_{\a}[\,H^n\,]^{\a}_{\b}=\Box_{\rm s}^n\nb_{\b}$ as well as any of its smooth operator-valued functions, $\nb_{\a}[\,f(H^n)\,]^{\a}_{\b}=f(\Box_{\rm s})\nb_{\b}$,

Therefore, the heat kernel (\ref{HKforF}) takes the form
\begin{equation}
\begin{split}
K^\mu_{\nu, F}(\tau)&= K^\mu_{\nu, H}(\tau)e^{-m^2\tau}
-e^{-\tau m^2}\nb^{\m}\frac{e^{\tau\Box_{\rm s}}-1}{\Box_{\rm s}}\nb_{\n}.
\end{split}
\label{HKforFnice}
\end{equation}
By a straightforward calculation one can check that this heat kernel satisfies the heat equation with appropriate initial conditions. This is one of the central results of this paper.

As we see, this heat kernel has a nontrivial  nonlocal structure involving the Green's function of $\Box_{\rm s}$ convoluted with its heat kernel, which is essentially different from that of the minimal operators. Rather than writing down its small $\tau$ expansion (which is the subject of future research) we go over directly to its functional trace.
\\

\subsection{Double pole divergences of the one-loop effective action}
The functional trace of $K^\mu_{\nu, F}(\tau)$ na\"{\i}vely reads as
\begin{equation}
\begin{split}
\Tr K^\mu_{\nu, F}(\tau)
&=\Tr K^\mu_{\nu, H-m^2}(\tau)-\Tr K_{\Box_{\rm s}-m^2}(\tau),
\end{split}
\end{equation}
where the second term arises by formally using in the second term of (\ref{HKforFnice}) the cyclic permutation under the trace sign and disregarding the $\delta(0)$ term, $\rm Tr\,1\propto \delta(0)$. However, under a closer inspection of the kernel of the operator in the second term of (\ref{HKforFnice})
\begin{equation}
\nb^{\m}\frac{e^{\tau\Box_{\rm s}}-1}{\Box_{\rm s}}\nb_{\n}\delta(x,y)=
-\nabla_\n^y\lb\,\nb^{\m}\frac{e^{\tau\Box_{\rm s}}-1}{\Box_{\rm s}}\delta(x,y)\,\rb
\end{equation}
this cyclic permutation involves integration by parts according to the relation $\nabla_\mu^y f^\mu(x,y)\,|_{\,y=x}=\nabla_\mu[f^\mu(x,x)]-\nabla_\mu f^\mu(x,y)\,|_{\,y=x}$, and we get for (\ref{CorrectGeomProgr1}) the expression which in contrast to~\cite{barvin1985} explicitly contains the total derivative term
\begin{equation}
\begin{split}
&\Tr \nb^{\m}\frac{e^{\tau\Box_{\rm{s}}}-1}{\Box_{\rm{s}}}\nb_{\m}
=\Tr\Big[\,K_{\Box_{\rm s}}(\tau)-1\Big]\\
&\qquad\quad-\int d^dx\nb_{\m}\lb\nb^{\m}
\frac{e^{\tau\Box_{\rm{s}}}-1}{\Box_{\rm{s}}}\d(x,y)\,\Big\vert_{\,y=x}\rb.
\end{split}
\label{HKtraceSurf}
\end{equation}
\\
Thus the heat kernel trace reads modulo $\propto\d(0)$-terms
\begin{equation}
\begin{split}
&\!\!\!\Tr K^\m_{F,\n}(\tau)=\Tr\Big[\, K^\m_{\n,H}(\tau)-\Tr K_{\Box_{\rm{s}}}(\tau)\Big]\,e^{-\tau m^2}\\
&\quad+\int d^dx\,\nb_{\m}\lb\nb^{\m}\frac{e^{\tau\Box_{\rm s}}-1}{\Box_{\rm s}}\d(x,y)\,\Big\vert_{\,y=x}\rb\,e^{-\tau m^2}.
\end{split}
\end{equation}
The latter term in (\ref{HKtraceSurf}) is a surface term and it will yield a double pole in the effective action.
Let us now see if our new heat kernel yields the same pole structure for $\Tr\ln F^{\m}_{\n}(m^2)$ as does the perturbation theory. The 1-loop effective action then reads
\begin{equation}
\begin{split}
&\frac12\Tr\ln F^{\m}_{\n}(m^2|\nabla)
=\frac12\Tr\ln \big[\,H^{\m}_{\n}(\nabla)-m^2\delta^\mu_\nu\big]\\
&\qquad\qquad\qquad\quad\;\;\;-\frac12\Tr\ln \big[\,\Box_{\rm s}-m^2\big]\\
&-\frac12\int\limits_0^{\infty}\frac{d\tau}{\tau}e^{-\tau m^2}\!
\int d^dx\,\nb_{\m}\lb\nb^{\m}\frac{e^{\tau\Box_{\rm s}}-1}{\Box_{\rm s}}\d(x,y)\,\Big\vert_{\,y=x}\rb.
\end{split}
\end{equation}
The first two terms correspond to $d-1$ physical degrees of freedom of the Proca field---$d$ vector modes minus one longitudinal degree of freedom in this theory with one second-class constraint. The additional total derivative term can be calculated by using the integral representation for the nonlocal operator $(e^{\tau\Box_{\rm s}}-1)/\Box_{\rm s}=\int_0^\tau ds\, e^{\tau\Box_{\rm s}}$ and substituting the Schwinger-DeWitt expansion (\ref{HeatKernel0}) for the heat kernel of $\Box_{\rm s}$. This leads to the following series
\begin{widetext}
\begin{equation}
\begin{split}
&-\frac12\int\limits_0^{\infty}\frac{d \tau}{\tau}e^{-\tau m^2}\int d^dx\,\nb_{\m}\lb\,\nb^{\m}\frac{e^{\tau\Box_{\rm s}}-1}{\Box_{\rm s}}\d(x,y)\,\Big\vert_{\,y=x}\rb
=-\frac12\int\limits_0^{\infty}\frac{d \tau}{\tau}\int\limits_0^{\tau}ds\int d^dx\,\nb_{\m}\lb\,\nb^{\m}\,e^{s\Box_{\rm s}-\tau m^2}\d(x,y)\,\Big\vert_{\,y=x}\rb\\
&\qquad\qquad\qquad=-\frac12\int \frac{d^dx\,g^{1/2}}{(4\pi)^{d/2}}\sum_{n=0}^{\infty}\lp m^2\rp^{-n+d/2 -1}\frac{ \G (n-d/2+1)}{n-d/2 +1}\nb_{\m}\lb\,\nb^{\m}a_{n}(\Box_{\rm s}|x,y)\,\Big\vert_{\,y=x}\rb,
\end{split}
\label{proca_hk_surf_tr}
\end{equation}
\end{widetext}
where $a_{n}(\Box_{\rm s}|x,y)$ are the HaMiDeW coefficients for scalar d'Alembertian $\Box_{\rm{s}}$. In any even-dimensional spacetime this series contains a double pole for $n=d/2-1$ (except the two dimensional case when $\nb^\m a_{0}(\Box_{\rm s}|x,y)\vert_{\,y=x}=0$). In four dimensions this yields the double pole divergence which, in view of the coincidence limit  $\nb^\m a_1(\Box_{\rm s}|x,y)\vert_{\,y=x}=\tfrac1{12}\nabla^\mu R$ \cite{barvin1985}, equals
\begin{equation}
\frac{1}{2}\Tr\ln F^{\m}_{\n}(m^2|\nabla)\Big\vert^{\rm{2-pole}}
\!\!\!\!\!\!=-\frac{1}{(\omega-2)^2} \int \frac{d^4x\,g^{1/2}}{32\pi^2}
\frac{1}{12}\Box R,
\label{double-pole}
\end{equation}
and thus coincides with the result (\ref{proca_pt_ans}) obtained by the method of generalized Schwinger-DeWitt technique in~\cite{barvin1985}.

\subsection{Boundary terms summation method}
\label{sec:ward}
There is another possible way to see how does this double-pole term arise. Following~\cite{barvin1985} we can make use of the Ward identity~(\ref{ward})
\begin{equation}
F^{\mu}_{\a}(m^2)(\d^{\a}_{\n}-\frac{1}{m^2}\nb^{\a}\nb_{\n})=(\Box-m^2)\d^{\m}_{\n}-R^{\m}_{\n}.
\label{WardIdProca}
\end{equation}
From (\ref{WardIdProca}) it follows that {modulo the anomaly},
\begin{equation}
\begin{split}
&\Tr\ln F^{\m}_{\n}(m^2)=\Tr\ln\lb (\Box-m^2)\d^{\m}_{\n}-R^{\m}_{\n} \rb\\
&\qquad\qquad-\Tr\ln\lb \d^{\m}_{\n}-\frac{1}{m^2}\nb^{\m}\nb_{\n} \rb+\d(0)(\dots),
\end{split}
\label{WardIdDetsProca}
\end{equation}
{assuming that the determinant anomaly, even if it is present here, should not contribute the double pole term to the one-loop functional determinant.}
The first term in r.h.s. of (\ref{WardIdDetsProca}) is  $\Tr\ln$ of a minimal operator, which can be calculated directly by the Schwinger--DeWitt technique. Its divergent part will contain only simple poles. The operator in the second term can also be transformed into a minimal one. Expanding the logarithm we obtain
\begin{equation}
\begin{split}
&\Tr\ln\lb\d^{\m}_{\n}-\frac{1}{m^2}\nb^{\m}\nb_{\n}\rb=-\Tr\sum_{n=1}^{\infty}\frac{1}{n}\nb^{\m}\frac{\Box_{\mathrm{s}}^{\,n-1}}{m^{2n}}\nb_{\n}\\
&\quad=\Tr\nb^{\m}\frac{1}{\Box_{\rm s}}\ln\lp1-\frac{\Box_{\mathrm{s}}}{m^{2}}\rp\nb_{\n}\\
&\quad=-\int d^dx\,\nb_\m^y\Big[\nb^{\m}\frac{\ln(1-\Box_{\mathrm{s}}/m^2)}{\Box_{\mathrm{s}}}\d(x,y)\Big]\,\Big|_{\,y=x}.
\end{split}
\label{CorrectGeomProgr1}
\end{equation}
Here it is understood that the kernel of the operator $\delta^\mu_\nu-\nabla^\mu\nabla_\nu/m^2$ should be derived from its action on a vector test function $\varphi^\nu(y)$ which admits integration by parts in view of its compact support. Thus any $n$-th power of $\nabla^\mu\nabla_\nu$ acting on $\varphi^\nu(x)$ gives rise to the $(n-1)$-th power of the scalar d'Alembertian $\Box_{\rm s}$,
\begin{equation}
\nabla^\mu\Box^{n-1}_{\rm s}\nabla_\nu\varphi^\nu(x)=-\int d^dx\,\nabla_\nu^y\big[\nabla^\mu_x\Box^{n-1}_{\rm s}\delta(x,y)\,\big]\varphi^\nu(y),
\end{equation}
whence after the summation over $n$ we get (\ref{CorrectGeomProgr1}) with the derivatives acting on different arguments $x$ and $y$ before their identification in the trace operation. Thus, using the obvious integration by parts relation, $\nabla_\mu^y f^\mu(x,y)\,|_{\,y=x}=\nabla_\mu[f^\mu(x,x)]-\nabla_\mu f^\mu(x,y)\,|_{\,y=x}$ we get for (\ref{CorrectGeomProgr1}) the expression which in contrast to~\cite{barvin1985} explicitly contains the total derivative term
\\
\begin{widetext}
\begin{equation}
\begin{split}
\Tr\ln\lb\d^{\m}_{\n}-\frac{1}{m^2}\nb^{\m}\nb_{\n}\rb&=
\int d^dx\lc \Big[ \nb^{\m}\nb_{\m}\frac{\ln\lp1-\Box_{\mathrm{s}}/m^2\rp}{\Box_{\mathrm{s}}} \d(x,y)\Big]\,\Big\vert_{y=x}
-\nb_{\m}\Big[  \nb^{\m}\frac{\ln\lp1-\Box_{\mathrm{s}}/m^2\rp}{\Box_{\mathrm{s}}}\d(x,y)\Big\vert_{y=x}\Big]  \rc\\
&=\Tr\ln\lb1-\frac{\Box_{\mathrm{s}}}{m^2}\rb-\int d^dx \nb_{\m}\Big[  \nb^{\m}\frac{\ln\lp1-\Box_{\mathrm{s}}/m^2\rp}{\Box_{\mathrm{s}}}\d(x,y)\Big\vert_{y=x}\Big].
\end{split}
\label{proca_ward_ibp}
\end{equation}
The first term is the ``volume'' term, whereas the second is the total derivative one, and it will yield the double pole. We now use mass integral representation for the logarithm
\begin{equation}
    \ln\lp1-\frac{\Box_{\mathrm{s}}}{m^2}\rp=\int\limits_{m^2}^{\infty}d\mu^2\lp\frac{1}{\mu^2}-\frac{1}{\mu^2-\Box_{\mathrm{s}}}\rp.
    \label{mass_int_rep}
\end{equation}
And by employing proper-time integral representation for the resulting Green's function we can explicitly evaluate the total derivative contribution at $d=4$
\begin{equation}
\begin{split}
     &\Tr\ln\lb\delta^{\mu}_{\nu}-\frac{1}{m^2}\nabla^{\mu}\nabla_{\nu}\rb_{\mathrm{total-derivative}}\!\!\!\!=\int d^4x\, \nb^{\m}\bigg[\nb_{\m}
     \int\limits_{m^2}^{\infty}\frac{d\mu^2}{\mu^2}\int\limits_0^{\infty}ds\,e^{s(\Box_{\mathrm{s}}-\mu^2)}\d(x,y)\Big\vert_{y=x}\bigg].
\end{split}
\end{equation}

Now, we can apply the standard expansion for the heat kernel of the scalar d'Alembertian. Divergences, however, appear not only from the proper time integration, but also from the mass $\mu^2$-integral as well,
\begin{equation}
\begin{split}
    &\Tr\ln\lb\delta^{\mu}_{\nu}-\frac{1}{m^2}\nabla^{\mu}\nabla_{\nu}\rb_{\mathrm{tot\,der}}\!\!\!\!=\int\frac{d^4x\,g^{1/2}}{16\pi^2}
    \sum_{n=0}^{\infty}\,\int\limits_{m^2}^{\infty}\frac{d\mu^2}{\mu^2}\int\limits_0^{\infty}\frac{ds}{s^{d/2-n}}e^{-s\mu^2}
    \nb^{\m}\bigg[\nb_{\m}a_n^{\Box}(x,y)\Big\vert_{y=x}\bigg],
\end{split}
\end{equation}
where $a_n^{\Box}$ are the HaMiDeW coefficients for the scalar d'Alembert operator $\Box_{\mathrm{s}}$. This yields
\begin{equation}
\Tr\ln\lb\delta^{\mu}_{\nu}-\frac{1}{m^2}\nabla^{\mu}\nabla_{\nu}\rb_{\mathrm{tot\,der}}\!\!\!\!=\int \frac{d^4x \, g^{1/2}}{16\pi^2}\sum_{n=0}^{\infty}\frac{\Gamma(1+n-d/2)}{1+n-d/2}(m^2)^{-1-n+d/2}\nb_{\m}\lb\nb^{\m} a_n^{\Box}(x,y)\big\vert_{y=x}\rb.
\label{surf_sch-dw_expansion}
\end{equation}
\end{widetext}
As soon as $\nb_{\a}a_0^{\Box}(x,y)\big\vert_{y=x}=0$ and $\nb_{\a}a_1^{\Box}(x,y)\big\vert_{y=x}=\frac{1}{12}\nb_{\a}R$, the divergent part of the total derivative term sum in~(\ref{CorrectGeomProgr1}) reads
\begin{equation}
\begin{split}
&\Tr\ln\lb\delta^{\mu}_{\nu}-\frac{1}{m^2}\nabla^{\mu}\nabla_{\nu}\rb^{\mathrm{2-pole}}\\
&=\frac{1}{(\omega-2)^2}
\int \frac{d^4x\,g^{1/2}}{16\pi^2}\frac{\Box R}{12},\quad\omega\rightarrow2-0,
\end{split}
\label{proca_ward_double_pole}
\end{equation}
where the pole of order 2 is again due to superimposition of singularities in $\mu^2$ and $s$ integrals for $n=1$ term in Schwinger--DeWitt series.

Let us also note, that substituting~(\ref{proca_ward_double_pole}) into~(\ref{proca_ward_ibp}) and then into~(\ref{WardIdDetsProca}) yields the same answer as perturbation theory~(\ref{proca_pt_ans}), since the other terms are double-pole free.

\section{Determinant anomalies and surface terms}
\label{sec:det_anomalies}
In this section we will use the following notations. We consider a compact $d$-dimensional manifold $\mathcal{M}$ with a boundary $\pp \mathcal{M}$ and a vector bundle $\pi:\mathcal{E}\rightarrow\mathcal{M}$, which means that the differential operators are matrix valued. We denote spacetime indices on $\mathcal{M}$ with the letters of the Greek alphabet and use index-free notation in the bundle, so that the matrix nature of the operators will be denote by the hat, and the matrix trace over the spin-tensor indices $A$ of the bundle will be denoted by $\tr$, $\hat X\equiv X^A_B$, $\tr\hat X=X^A_A$.  Let the connection $\nb_{\a}$ on the sections of the bundle be torsionless and compatible with manifold metric $g_{\a\b}$. Let the associated curvatures of $\nb_{\a}$ acting on the spacetime vectors and the fibre bundle fields $\varphi\equiv\varphi^A$ be denoted as
\begin{equation}
[\nb_{\a},\nb_{\b}]\,v^{\g}=R^{\g}_{\phantom{\g}\a\b\l}v^{\l},\qquad[\nb_{\a},\nb_{\b}]\,\varphi=\h\calR_{\a\b}\,\varphi.
\end{equation}

Simple explanation why the determinant anomaly (\ref{anomaly_def}) should be given by a total-derivative term is based on the variation formula for the functional determinant. For the determinant $\Det\calO$  of the operator $\calO$ it is defined by the equation
\begin{equation}
\d\ln\Det\calO=\Tr\lb\d\calO\calO^{-1}\rb=\d\Tr\ln\calO,
\label{variation_def}
\end{equation}
where $\d\calO$ is a deformation of this operator which does not alter its order.
Let $\h\calO_{1}$, $\h \calO_{2}$ be elliptic differential operators with a composition $\h\calO_{12}=\h\calO_{1}\h\calO_{2}$ acting on the sections of vector bundle. From~(\ref{variation_def}), variation of the product operator determinant reads
\begin{equation}
\begin{split}
&\d\ln\Det[\h\calO_{12}]=\d[\Tr\ln \h\calO_{12}]=\Tr[\h\calO_{12}^{-1}\d\h\calO_{12}]\\
&=\Tr\lb\h\calO_2^{-1}\h\calO_1^{-1}\lp \d\h\calO_1\h\calO_2+\h\calO_1\d\h\calO_2 \rp\rb,
\end{split}
\label{variation}
\end{equation}
If we na\"{\i}vely apply trace cyclic permutation rule to~(\ref{variation}), we will disregard the fact that it involves integration by parts, which, in the case of a manifold with a boundary, produces a non-zero surface term via Stokes theorem. Thus,
\begin{equation}
\begin{split}
\d\ln\Det[\h\calO_{1}\h\calO_{2}]&=\d\ln\Det\h\calO_1+\d\ln\Det\h\calO_2\\
&+\int_{\cal M}d^dx\,\partial_\mu \lb\dots\rb^{\m},
\end{split}
\label{anomaly_naive}
\end{equation}
with ellipses denoting some total-derivative term.

Although, as we mentioned in Introduction, there exists a technique of calculating determinant anomalies (\ref{anomaly_def}) with the help of the Wodzicki residue~\cite{kontsevich1994, wodz_book}, we do not employ it. Our goal is to examine the structure of the terms arising from integration by parts on the examples of generic operators of the second and fourth orders. In what follows we provide calculations of divergent parts for several determinant anomalies in order to illustrate that they indeed have the form of integrals of total-derivative terms.

For these calculations we employ the Schwinger--DeWitt technique~\cite{gilkey_orig, schwinger_orig, dewitt_book, schwinger_orig, barvin1985}, from which it follows that the divergent in even $d$ dimensions (i.e. at $\omega\rightarrow d/2-0$) part of the determinant logarithms for the second and the fourth order operators $\h F_{(2)}$, $\h F_{(4)}$ read respectively
\begin{equation}
\ln\Det \h F_{(2)}\big\vert^{\rm div}=\frac{1}{\omega-d/2}\int d^dx\,g^{1/2}\tr \h E_d^{\h F_{(2)}}(x),
\label{det_2_div}
\end{equation}
\begin{equation}
\ln\Det \h F_{(4)}\big\vert^{\rm div}=\frac{2}{\omega-d/2}\int d^dx\,g^{1/2}\tr\h E^{\h F_{(4)}}_d(x),
\label{det_4_div}
\end{equation}
where $\h E_{2m}^{\h F_{(2)}}$ and $\h E_{2m}^{\h F_{(4)}}$ are respectively the Gilkey--Seeley coefficients for $\h F_{(2)}$ and $\h F_{(4)}$ ($\h E_{2m}(x)=(4\pi)^{-d/2}\h a_{m}(x,y)\,\vert_{\,y=x}$) participating in the early-time ($s\rightarrow0$) asymptotic expansions of their heat kernel coincidence limits
\begin{equation}
e^{\tau \h F_{(2)}}\delta(x,y)\big\vert_{y=x}=\frac{g^{1/2}(x)}{\tau^{d/2}}\sum_{m=0}^{\infty}\tau^m\h E^{\h F_{(2)}}_{2m}(x),
\end{equation}
\begin{equation}
e^{\tau \h F_{(4)}}\delta(x,y)\big\vert_{y=x}=\frac{g^{1/2}(x)}{\tau^{d/4}}\sum_{m=0}^{\infty}\tau^{m/2}\h E^{\h F_{(4)}}_{2m}(x).
\end{equation}
From~(\ref{det_2_div}---\ref{det_4_div}) it follows that the divergent part of the anomaly~(\ref{anomaly_def}) for second order operators $\h F_1$ and $\h F_2$ in $d=2,4$ reads
\begin{equation}
\begin{split}
&\mathcal{A}_{12}^{d\rightarrow2,4}\Big\vert^{\mathrm{div}}_{d}
=\frac{1}{\omega-d/2}\int d^dx\,g^{1/2}\\
&\times\tr\lb 2\h E_d^{\h F_{12}}-\h E_d^{\h F_1}-\h E_{d}^{\h F_2} \rb,\qquad\omega\rightarrow d/2-0,
\end{split}
\label{anom_E}
\end{equation}
where $\h F_{12}=\h F_1\h F_2$. Note, that since the traces of $\h E_{2m}$ in the formula above have an explicit dependence on dimension $d$, the concrete limits $\lim_{d\rightarrow2}\h E_{2}$ and $\lim_{d\rightarrow4}\h E_{4}$ should be separately taken.

Calculations below have been performed with the help of {\rm Mathematica} package \textit{xAct}~\cite{xact}.
\\
\\

\subsection{Determinant anomaly for minimal operators}
\label{sec:anomaly_minimal}
We define the operators:
\begin{equation}
\hat{{F}}_{1}=\Box\hat{1}+\hat{A}_{\alpha}\nb^{\alpha}+\hat{Q},\qquad\hat{{F}}_{2}=\Box\hat{1}+\hat{P},
\label{minops_for_anom}
\end{equation}
and calculate with the help of~(\ref{anom_E}) the divergent part of their anomaly~(\ref{anomaly_def}) in 2 and 4 dimensions.

The coefficients $\h E_{2}$ and $\h E_4$ for a general minimal second order operator $\hat{{F}}_{2}$
read~\cite{barvin1985}
\begin{equation}
\begin{split}
&\tr\h E_2^{\h F_2}=\frac{1}{4\pi}\bigg[ P+\frac{\tr\h1}{6}R\bigg],\\
&\tr\h E_4^{\h F_2}=\frac{1}{16\pi^2}\bigg[ \frac{\tr \h 1}{180}( R^2_{\a\b\m\n}-R^2_{\a\b}+\frac{5}{2}R^2)\\
&\qquad\qquad+\frac{\tr \h 1}{30}\,\Box R+\frac{1}{12}\tr\h{\cal R}_{\a\b}^2\\
&\qquad\qquad+\frac{1}{2}\tr\h P^2+\frac{1}{6}PR+\frac{1}{6}\Box P \bigg].
\end{split}
\label{trlogF2_d4}
\end{equation}

These algorithms can also be applied to $\h F_{1}$ by absorbing the term $\h A^{\a}\nb_{\a}$ linear in derivatives in the redefinition of covariant derivatives. That is, we shift the covariant derivative $\nb_{\a}$, which alters the curvature $\h{\cal R}_{\a\b}$:
\begin{equation}
\begin{split}
\nb_{\a}&\mapsto{\t\nb}_{\a}=\nb_{\a}+\frac{1}{2}\h A_{\a},\\
\h\calR_{\a\b}&\mapsto\h{\cal R}_{\a\b}+\nb_{[\a}\h A_{\b]}+\frac{1}{2}\h A_{[\a}\h A_{\b]}.
\end{split}
\end{equation}
After this shift, the operator $\h{F}_1$ reads
\begin{equation}
\begin{split}
\h{F}_1&=\t \Box-\frac{1}{2}\t\nb_{\a}\h A^{\a}-\frac{1}{4}\h A_{\a}^2+\h Q,
\end{split}
\end{equation}
so that
\begin{equation}
\!\!\!\tr\h{E}_{2}^{\h F_1}=\frac{1}{4\pi}\tr\lb \frac{\h1}{6}R+\h Q-\frac{1}{2}\nb_{\a}\h A^{\a}-\frac{1}{4}\h A_{\a}\h A^{\a} \rb,
\label{det_F1_d2}
\end{equation}
\begin{widetext}
\begin{equation}
\begin{split}
\!\!\!\!\tr\h{E}_{4}^{\h F_1}&=\frac1{16\pi^2}\tr\bigg[\frac{\h1}{180}\lp R^2_{\a\b\m\n}-R^2_{\a\b}
+\frac{5}{2}R^2\rp+\frac{\h1}{30}\Box R+\frac{1}{12}\h\calR^2_{\a\b}+\frac{1}{2}\h Q^2+\frac{1}{6}\h QR
+\frac{1}{6}\Box\h Q\\
&-\frac{1}{2}\nb_{\a}\h A^{\a}(\h Q+\frac{\h1}{6}R)+\frac{1}{6}\h\calR_{\a\b}\nb^{\a}\h A^{\b}-\frac{1}{12}\Box\nb_{\a}\h A^{\a}
-\frac{1}{4}\h A_{\a}^2(\h Q+\frac{\h 1}{6}R)+\frac{1}{12}\h\calR_{\a\b}\h A^{\a}\h A^{\b}-\frac{1}{12}\h A_{\a}\Box\h A^{\a}\\
&+\frac{1}{8}(\nb_{\a}\h A^{\a})^2-\frac{1}{12}(\nb_{(\a}\h A_{\b)})^2\!+\frac{1}{12}\h A^{\a}\h A^{\b}\nb_{[\a}\h A_{\b]}\!+\frac{1}{8}\h A^2_{\a}\nb_{\b}\h A^{\b}\!+\frac{1}{48}\h A^2_{\a}\h A_{\b}^2\!+\frac{1}{96}(\h A_{\a}\h A_{\b})^2\bigg].
\end{split}
\label{det_F1_d4}
\end{equation}
\end{widetext}
The product operator $\h F_{12}$ reads
\begin{equation}
\begin{split}
&\hat{F}_{12}=\h F_1\h F_2=\Box^2\hat{1}+\hat\O_{(\a\b\g)}\nb^{\a}\nb^{\b}\nb^{\g}\\
&+\hat D_{(\a\b)}\nb^{\a}\nb^{\b}+\hat{H}_{\a}\nb^{\a}+\hat{U},
\end{split}
\label{4thorder_operator}
\end{equation}
with the coefficients (we have symmetrized~$\hat\O_{\a\b\g}$ over spacetime indices)
\begin{equation}
\hat{\O}_{\a\b\g}=\frac{1}{3}\lp g_{\b\g}\hat{A}_{\a}+g_{\g\a}\hat{A}_{\b}+ g_{\a\b}\hat{A}_{\g}\rp,
\end{equation}
\begin{equation}
\hat D_{\a\b}=\lp\hat{P}+\hat{Q}\rp g_{\a\b},
\end{equation}
\begin{equation}
\hat{H}_{\a}=-\frac{1}{3}\hat{A}^{\b}\lp 2R_{\a\b}+3\,\hat{\mathcal{R}}_{\a\b}\rp+2\nb_{\a}\hat{P}+\hat{A}_{\a}\hat{P},
\end{equation}
\begin{equation}
\hat{U}=\frac{1}{3}\hat{A}_{\a}\nb_{\b}\hat{\mathcal{R}}^{\a\b}+\hat{A}_{\a}\nb^{\a}\hat{P}+\Box\hat{P}+\hat{Q}\hat{P}.
\end{equation}

Using the formulae for the fourth order operator, given in~\cite{gus_4_order} and~\cite{4thorder_22}, we obtain~$\tr\h E_2^{F_{12}}$ and the  $\O$-independent part of~$\tr\h E_4^{F_{12}}$ in an arbitrary dimension $d$
\begin{widetext}
\begin{equation}
\begin{split}
&\tr\h E_2^{\h F_{12}}=\frac{1}{(4\pi)^{d/2}}\frac{\G\lp \frac{d-2}{4} \rp}{\G\lp \frac{d}{2}-1 \rp}\bigg[ -\frac{d+2}{32d}\tr\h A^2_{\a}+\frac{1}{4}(P+Q)+\frac{\tr\h1}{12}R-\frac{d+2}{8d}\nb_{\a}A^{\a} \bigg],\\
&\tr \h E_4^{\h F_{12}}\Big\vert_{\O=0}
=\frac{1}{(4\pi)^{d/2}} \frac{\Gamma\left(\frac{d}{4}\right)}{4\Gamma\left(\frac{d}{2}\right)}
\Bigg[(d-2)\bigg(\frac{\tr\hat 1}{90}R^2_{\a\b\m\n}-
\frac{\tr\hat 1}{90}R^2_{\a\b}+\frac{\tr\hat 1}{36} R^2 + \frac{1}{6} \hat{\mathcal{R}}^2_{\a\b} + \frac{\tr\hat 1}{15}\,\Box R\bigg)\\
&\qquad\qquad+\frac{d-2}{6}(P+Q)R+\frac{d}{4}\tr(\h P+\h Q)^2+\frac{d-2}{6}\,\Box (P+Q)-2\tr\h Q\h P\\
&\qquad\qquad-\tr\h A^{\a}\nb_{\a}(\h P+\frac{\h1}{3}R)+\frac{1}{3}\tr\h A_{\a}\nb_{\b}\h{\cal R}^{\a\b}-\frac{2}{3}\nb_{\a} A_{\b}R^{\a\b}-\tr\nb_{\a}\h A_{\b}\h{\cal R}^{\a\b}{+\tr\nb_{\a}\h A^{\a}\h P}
\Bigg].
\end{split}
\end{equation}
\end{widetext}
The $\O$-dependent part of $\h E_4$ of a generic minimal fourth order differential operator can be represented by the sum of the terms of Eq.(\ref{A1}) in Appendix \ref{sec:appendix_a}. Their derivation is based on the corrected algorithm for a minimal fourth-order operator with a generic third order term in derivatives~\cite{4thorder_22,BKW}.\footnote{Modulo several typos these contributions have been first calculated in~\cite{4thorder_22}. Corrected expressions---corrections affect terms $\h B_{\O,2}$, $\h B_{\O,3}$, $\h B_{\O R}$, and $\h B_{\O\calR}$---are presented in~\cite{BKW}.} Using them we obtain $\O$-dependent contributions into $\tr\h E_4^{\h F_{12}}$, which, due to their complexity are listed in \hyperref[sec:appendix_a]{Appendix A}.

Let us emphasize once again, that we should not compare coefficients $\h E_{2m}^{F_{21}}$ with the sum of those of $\h F_{1,2}$ in an arbitrary dimension $d$, since these coefficients are ``unphysical'' by themselves. We anticipate them to coincide (modulo total derivative terms) in view of~(\ref{anomaly_naive}) only if $2m=d$, since only in this case they represent the divergent part of the 1-loop effective action. Therefore we calculate the quantity~(\ref{anom_E}) separately in $d=2$ ($\omega\rightarrow1-0$) and $d=4$ ($\omega\rightarrow2-0$), which, after some transformations turn out to be, respectively
\begin{equation}
\mathcal{A}^{d\rightarrow2}_{12}\big|^{\rm div}=-\frac{1}{\omega-1}\int \frac{d^2x\,g^{1/2}}{8\pi}\nb_{\a}A^{\a},
\label{minimal_d=2_anomaly}
\end{equation}
\begin{widetext}
\begin{equation}
\begin{split}
\mathcal{A}_{12}^{d\rightarrow4}\big\vert^{\rm div}&=\frac{1}{\omega-2}\int \frac{d^4x\,g^{1/2}}{16\pi^2}\nb_{\a}\tr\bigg[-\frac{1}{4}\h A^{\a}(\h P+\h Q)-\frac{1}{12}(\h A^{\a}R+\h A_{\b}R^{\a\b})-\frac{1}{9}\nb^{\a}\nb_{\b}\h A^{\b}-\frac{7}{36}\Box \h A^{\a}
\\&+\frac{2}{9}\nb_{\b}\nb^{\a}\h A^{\b}+\frac{11}{72}\h A^{\a}\nb_{\b}\h A^{\b}-\frac{1}{72}\lp \nb^{\a}\h A^{\b}\h A_{\b}+\nb_{\b}\h A^{\a}\h A^{\b} \rp+\frac{1}{24}\h A^{\a}\h A_{\b}\h A^\beta\bigg],\qquad\omega\rightarrow2-0.
\end{split}
\label{minimal_d=4_anomaly}
\end{equation}
\end{widetext}
As one can see, in both $d=2$ and $d=4$ cases the divergent part of the determinant anomaly $\cal A_{12}$ is proportional to the integral of total-derivative terms, which confirms conclusions drawn from~(\ref{anomaly_naive}).

\subsection{Determinant anomaly for nonminimal operators}
\label{sec:anomaly_nonminimal}
To check the determinant anomaly for nonminimal operators is not easy, because the algorithm for nonminimal fourth-order operator is unknown and the product of two nonminimal operators is generally also nonminimal. However, there is an exceptional case when the product of two second-order nonminimal operators is a minimal one~\cite{Shapiro}. When the parameters $\vk$ and $\l$ of the following two operators
\begin{equation}
\begin{split}
F_1{}^{\a}_{\b}(\vk)&=\Box\d^{\a}_{\b}-\vk\nb^{\a}\nb_{\b}+X^{\a}_{\b},\\
F_2{}^{\a}_{\b}(\l)&=\Box\d^{\a}_{\b}-\l\nb^{\a}\nb_{\b}+Y^{\a}_{\b},
\end{split}
\end{equation}
are related by the equation $\vk=\frac{\l}{\l-1}$, the product operator is minimal,
\begin{equation}
\begin{split}
&F_{12}{}^{\a}_{\b}=F_1{}^{\a}_{\g}(\l)F_2{}^{\g}_{\b}(\vk)\\
&=\Box^2\d^{\a}_{\b}+D^{\a}_{\b}{}^{\m\n}\nb_{\m}\nb_{\n}+H^{\a}_{\b}{}^{\m}+U^{\a}_{\b},
\end{split}
\end{equation}
with the tensor coefficients
\begin{equation}
\begin{split}
&D^{\a}_{\b}{}^{\m\n}=(X^{\a}_{\b}+Y^{\a}_{\b})g^{\m\n}\\
&\qquad\qquad-\frac{\vk}{2}\lb (Y^{\m}_{\b}+R^{\m}_{\b})g^{\a\n}+(Y^{\n}_{\b}+R^{\n}_{\b})g^{\a\m} \rb\\
&\qquad\qquad-\frac{\l}{2}\lb (X^{\a\m}+R^{\a\m})\d^{\n}_{\b}+(X^{\a\n}+R^{\a\n})\d^{\m}_{\b} \rb,\\
\end{split}
\end{equation}
\begin{equation}
\begin{split}
&H^{\a}_{\b}{}^{\m}=-\vk\nb^{\a}(Y^{\m}_{\b}+R^{\m}_{\b})\\
&\qquad\qquad-\vk\lp\frac{1}{2}\nb_{\b}R+\nb_{\g}Y^{\g}_{\b}\rp g^{\m\a}+2\nb^{\m}Y^{\a}_{\b},\\
&U^{\a}_{\b}=\vk R^{\a}_{\g}(R^{\g}_{\b}+Y^{\g}_{\b})-\vk\nb_{\g}\nb^{\a}(R^{\g}_{\b}+Y^{\g}_{\b})\\
&\qquad\quad+X^{\a}_{\g}Y^{\g}_{\b}+\Box Y^{\a}_{\b}-\frac{\vk}{2}(Y^{\m\g}+R^{\m\g})R_{\g\b\m}^{\phantom{\g\b\m}\a}\\
&\qquad\quad\qquad\qquad+\frac{\l}{2}(X^{\a\m}+R^{\a\m})R_{\b\m},
\end{split}
\end{equation}
where everywhere $\vk=\l/(\l-1)$, and we have symmetrized $D^{\a}_{\b}{}^{\m\n}$ over the last two indices.

Using~(\ref{anom_E}), we evaluate the anomaly. The contribution of $\tr\h E_4^{F_{12}}$ is calculated by using the same coefficients as for the fourth order operator in the previous section on minimal operators anomaly. In the case of general matrices $X^{\a}_{\b}$ and $Y^{\a}_{\b}$ there are 19 tensor structures in the total $\tr\h E_4$ which at $d\rightarrow4$ reads
\begin{equation}
\begin{split}
&\tr \h E^{F_{12}}_4=\frac{1}{16\pi^{2}}\bigg[\co{1}R^2_{\a\b}+\co{2}R^2+\co{3}R^2_{\a\b\m\n}\\
&+\co{4}RX+\co{5}R_{\a\b}X^{\a\b}+\co{6}X^2_{\a\b}+\co{7}X^{\a\b}X_{\b\a}\\
&+\co{8}X^2+\co{9}RY+\co{10}R_{\a\b}Y^{\a\b}+\co{11}Y^2_{\a\b}\\
&+\co{12}X^{\a\b}Y_{\b\a}+\co{13}Y^{\a\b}Y_{\b\a}+\co{14}Y^2+\co{15}\Box R\\
&+\co{16}\nb_{\a}\nb_{\b}X^{\a\b}+\co{17}\nb_{\a}\nb_{\b}Y^{\a\b}\\
&+\co{18}\Box X+\co{19}\Box Y\bigg],
\end{split}
\label{E4_4order_nonmin}
\end{equation}
with the numerical coefficients $\co{1}$--$\co{19}$ given in~\hyperref[sec:appendix_b]{Appendix B}.

The coefficients $\tr\h E_4^{F_{1}}$ and $\tr\h E_4^{F_{2}}$ can be calculated at $d=4$ by using the coefficients given in~\cite{gus_nonmin} and~\cite{Shapiro}.

With $\l\mapsto\l/(\l-1)$ for $\h F_2$ we evaluate the nonminimal operators anomaly in $d=4$ (as $\omega\rightarrow2-0$). It reads
\begin{equation}
\begin{split}
&\mathcal{A}_{12}^{d\rightarrow4}\bigg\vert^{\rm{div}}=\frac{1}{\omega-2}\int\frac{ d^4x \,g^{1/2}}{16\pi^2}\\
&\times\bigg\{ \frac{(5 \lambda -4) ((\lambda -2) \lambda +2 (\lambda -1) \ln
   (1-\lambda ))}{12 \lambda ^2}\nb_{\a}\nb_{\b}X^{\a\b}
   \\&+\frac{(\lambda +1) ((\lambda -2) \lambda +2 (\lambda -1) \ln (1-\lambda
   ))}{12 \lambda ^2}\Box X
\\&-\frac{(\lambda +4) ((\lambda -2) \lambda +2 (\lambda -1) \ln (1-\lambda
   ))}{12 (\lambda -1) \lambda ^2}\nb_{\a}\nb_{\b}Y^{\a\b}
   \\&-\frac{(2 \lambda -1) ((\lambda -2) \lambda +2 (\lambda -1) \ln
   (1-\lambda ))}{12 (\lambda -1) \lambda ^2}\Box Y
\\&+\frac{7 (\lambda -2) ((\lambda -2) \lambda +2 (\lambda -1) \ln
   (1-\lambda ))}{24 (\lambda -1) \lambda }\Box R\bigg\}.
\end{split}
\label{nonmin_anom_d4}
\end{equation}

As expected, this is a sum of total derivative term. Quite interestingly, this contribution identically vanishes in the case of $X^{\a\b}=Y^{\a\b}=-R^{\a\b}$. Apparently, this property can be attributed to the fact that in this case both operators $F_1$ and $F_2$ satisfy the Ward identity (\ref{ward}) which establishes special kind of relations between the coefficients of various structures in the final answer. This requires further analysis.

The two-dimensional case is much simpler. We have
\begin{equation}
\begin{split}
\tr \h E_2^{F_{12}}&=\frac{1}{4\pi}\bigg[ \frac{1}{4} (2-\lambda ) X+\frac{(\lambda -2) Y}{4 (\lambda-1)}\\
&+\frac{\left(3 \lambda ^2-4 \lambda +4\right)R}{12-12 \lambda } \bigg],\qquad d=2.
\end{split}
\label{1000}
\end{equation}

On the other hand from~\cite{gus_nonmin} at $d=2$ it follows that
\begin{equation}
\begin{split}
\tr\h E_2^{F_1}+\tr\h E_2^{F_2}&=\frac{1}{4\pi}\bigg[\left(1-\frac{\lambda }{2}\right) X+\frac{1}{2}\frac{\lambda -2}{\lambda -1}Y\\
&+\frac{3 \lambda ^2-4 \lambda +4}{6-6 \lambda }R\bigg],\qquad d=2.
\end{split}
\label{1001}
\end{equation}
Consequently, in view of (\ref{1000})~and~(\ref{1001}) at $d=2$ the divergent part of the anomaly is exactly zero,
\begin{equation}
\mathcal{A}_{12}^{d\rightarrow2}\big\vert^{\rm{div}}=0.
\label{nonmin_anom_d2}
\end{equation}

Thus, once again the anomaly both at $d=2$~(\ref{nonmin_anom_d2}) and $d=4$~(\ref{nonmin_anom_d4}) consists only of total-derivative terms, which, again confirms our conclusions drawn from~(\ref{anomaly_naive}).

\section{Conclusions}
Unexpected origin of double-pole terms in dimensionally regulated one-loop effective action of the Proca model in curved spacetime is explained by a nontrivial structure of the heat kernel for the nonminimal Proca operator. We derived this heat kernel by expressing it in terms of the heat kernels of the minimal vector field and the scalar d'Alembertian operators, the latter forming a special nonlocal convolution with the Green's function of the covariant scalar d'Alembertian (see Eq.(\ref{HKforFnice}) -- one of the main results of this paper). The nontrivial structure of the heat kernel is responsible for double-pole one-loop divergences which have the form of the total derivative contribution originating from integration by parts. Apparent contradiction between this result and the Gilkey-Seeley theory of heat kernel asymptotics for elliptic operators, which claims that only single pole divergences can be generated at one-loop order of semiclassical expansion, can be resolved by the observation that contrary to the assumptions of Gilkey-Seeley theorems the nonminimal Proca operator has a degenerate principal symbol. The inclusion of the mass term into this symbol, as a means of developing a systematic perturbation theory for the Proca Green's function and heat kernel, changes the typical structure of the heat kernel trace expansion and makes it consistent with its conventional version.

Another aspect of total-derivative terms in the quantum effective action, that was discussed in the paper, concerns the problem of multiplicative determinant anomalies---lack of factorization of the functional determinant of a product of operators into the product of their individual determinants. We gave a general argument in favor of the fact that this anomaly should have the form of total-derivative terms and checked this statement by considering several rather nontrivial examples by calculating the divergent parts of one-loop functional determinants for products of minimal and nonminimal second-order differential operators in curved spacetime. This result looks important because a widely recognized method of renormalization analysis for higher derivative theories (usually disregarding the renormalization of surface terms), like renormalizable quadratic gravity~\cite{Tseytlin,barvin1985,Avramidi-Barvinsky,buchbinder_book,Salvio-Strumia,Shapiro} and many others, is just based on the multiplication property of the functional determinants---decomposition of the higher-derivative inverse propagator into the product of second order differential factors and summation of their individual contributions to the full effective action.

Apart from the above justification, the obtained bulk total-derivative contributions to UV divergences thus far have only structural significance, because in concrete problems with spacetime boundaries there will be other surface contributions coming from the $B_n(x)$-terms of (\ref{A_n}), some of them being critically depending on boundary conditions for the operators in question, some of them---independent of those. Beyond their control within concrete boundary value setup their discussion becomes incomplete. In scattering problems on the background of asymptotically-flat spacetime most of these surface terms would be vanishing and insignificant, but in the modern context of cosmological setup or in the holographic AdS/CFT setup with asymptotically nontrivial boundaries and boundary conditions they become very important.

\section*{Acknowledgements}
Authors are indebted to {I. Shapiro, A. Kurov, and W. Wachowski for helpful comments on the present paper and fruitful discussions. We are also grateful to anonymous Referee for drawing our attention to the generation of total derivative terms by the use of a smearing function in the functional heat kernel trace}. This work was supported by the Russian Science Foundation Grant No. 23-12-00051, \url{https://rscf.ru/en/project/23-12-00051/}.

\appendix
\begin{widetext}
\section{Details of calculations of the minimal determinant anomaly}
\label{sec:appendix_a}
The $\O$-dependent part of $\tr\h E_4^{\h{F}_{12}}$ obtained in \cite{4thorder_22} and corrected in \cite{BKW} reads in $d$ dimensions as
\begin{equation}\label{A1}
\begin{split}
\tr\h E_4^{\h F_{12}}\Big\vert_\varOmega &= \frac{1}{(4\pi)^{d/2}} \frac{\Gamma\left(\frac{d}{4}\right)}{2\Gamma\left(\frac{d}{2}\right)} \tr\Big[ \hat B_{\varOmega,1}+\hat B_{\varOmega,2}
+\hat B_{\varOmega,3} + \hat B_{\varOmega R} + \hat B_{\varOmega\cal R} + \hat B_{\varOmega DH}\Big].
\end{split}
\end{equation}
The calculation of various contributions in the right hand side of this expression for our operator (\ref{4thorder_operator}) leads to
\begin{equation}
\begin{split}
&\tr\h B_{\O,1}^{\h F_{12}}=-\frac{d^2+2d +4}{24 (d +2)}\nb_{\a}\Box A^{\a}+\frac{d (d+4)}{24 (d+2)}\nb_{\a}\nb_{\b}\nb^{\a} A^{\b}-\frac{d^2+2 d+4}{24 (d+2)}\Box\nb_{\a} A^{\a},\\
&\tr\h B_{\O,2}^{\h F_{12}}=\frac{d+1}{12(d+2)}\tr\h A^{\a}\nb_{\a}\nb_{\b}\h A^{\b}+\frac{(d+4)(3d+8)}{96(d+2)}\tr(\nb_{\a}\h A^{\a})^2+\frac{d+1}{12(d+2)}\tr\h A^{\a}\nb_{\b}\nb_{\a}\h A^{\b}\\
&\qquad\qquad\quad-\frac{d^{2}+2d+4}{48(d+2)}\tr\h A^{\a}\Box\h A_{\a}-\frac{d(d+4)}{96(d+2)}\tr\nb_{\a}\h A_{\b}\nb^{\b}\h A^{\a}-\frac{d(d+4)}{96(d+2)}\tr(\nb_{\a}\h A_{\b})^2,\\
&\tr\h B_{\O,3}^{\h F_{12}}=\frac{(d+4)(d+8)}{1536(d+2)}\tr(\h A_{\a}\h A_{\b})^2+\frac{(d+4)(d+8)}{768(d+2)}\tr(\h A^{\a}\h A_{\a})^2+\frac{(d+4)(d+8)}{192(d+2)}\tr\h A^{\a}\h A^{\b}\nb_{\a}\h A_{\b}\\
&\qquad\qquad\quad+\frac{{(d-4)}(d+4)}{192(d+2)}\tr\h A^{\a}\h A^{\b}\nb_{\b}\h A_{\a}+\frac{(d+4)^2}{64(d+2)}\tr\h A_{\a}^2\nb_{\b}\h A^{\b},
\end{split}
\end{equation}
\begin{equation}
\tr \h B_{\O R}^{\h F_{12}}=\frac{1}{24} A^{\a}\nb_{\a}R+\frac{1}{4}\nb_{\a}A_{\b}R^{\a\b}
-\frac{d}{24}\nb_{\a}A^{\a}R+\frac{1}{12}\tr\h A^{\a}\h A^{\b}R_{\a\b}-\frac{d}{96}\tr\h A_{\a}^2 R,\\
\end{equation}
\begin{equation}
\tr\h B_{\O\calR}^{\h F_{12}}=-\frac{1}{6}\tr\h A^{\a}\nb^{\b}\h{\cal R}_{\a\b}
-\frac{d+4}{12}\tr\h{\cal R}_{\a\b}\nb^{\b}\h A^{\a}+\frac{d+4}{48}\tr\h A^{\a}\h A^{\b}\h{\cal R}_{\a\b},\\
\end{equation}
\begin{equation}
\begin{split}
&\tr\h B_{\O D H}^{\h F_{12}}
=-\frac{1}{6}\tr\h A_{\a}\h A_{\b} R^{\a\b}-\frac{1}{4}\tr\h A_{\a}\h A_{\b} \h\calR^{\a\b}+{\frac{4-d}{32}}\tr\h A^{2}_{\a}\h P-\frac{4+d}{32}\tr\h A^{2}_{\a}\h Q\\
&\qquad\qquad\quad+\frac{1}{4}\tr\h A^{\a}\nb_{\a}(\h P-\h Q)-\frac{d+2}{8}\tr\nb_{\a}\h A^{\a}(\h P+\h Q).
\end{split}
\end{equation}
\end{widetext}

\section{The coefficients of Eq.(\ref{E4_4order_nonmin})}
\label{sec:appendix_b}
\begin{eqnarray*}
&&\co{1}=\frac{15 \lambda ^4-60 \lambda ^3+44 \lambda ^2+32 \lambda
   -16}{720 (\lambda -1)^2},\nonumber\\
&&\co{2}=\frac{1}{144}
   \left(\frac{\left(3 \lambda
   ^2-10 \lambda -8 (\lambda
   -1)+10\right) \lambda ^2}{2
   (\lambda -1)^2}+8\right),\nonumber\\
\end{eqnarray*}
\begin{eqnarray*}
   &&\co{3}=-\frac{11}{180},\;\;
\co{4}=\frac{1}{24} \left(\frac{\lambda ^2}{2}-\lambda +2\right),\nonumber\\
&&\co{5}=-\frac{1}{72} (12-3 \lambda ) \lambda,\nonumber\\
&&\co{6}=\frac{\lambda ^2}{96},\;\;\co{7}=\frac{1}{96} \left(\lambda ^2-12
   \lambda +24\right),\;\;\co{8}=\frac{\lambda ^2}{96},\nonumber\\
&&\co{9}=\frac{1}{24} \left(\frac{\lambda ^2}{2 (\lambda
   -1)^2}-\frac{\lambda }{\lambda -1}+2\right),\nonumber\\
\end{eqnarray*}
\begin{eqnarray}
&&\co{10}=-\frac{\lambda  (8 (\lambda -1)+\lambda -4)}{72 (\lambda
   -1)^2},\;\;\co{11}=\frac{\lambda ^2}{96 (\lambda
   -1)^2},\nonumber\\
&&\co{12}=0,\;\;\co{13}=\frac{16 (\lambda -1)^2-8
   (\lambda -1)+(4-3 \lambda )
   \lambda }{96 (\lambda -1)^2},\nonumber\\
&&\co{14}=\frac{\lambda ^2}{96 (\lambda
   -1)^2},\;\;\co{15}=\frac{5 \lambda ^2+48 (\lambda -1)}{360 (\lambda -1)},\nonumber\\
&&\co{16}=\frac{5 \lambda }{36},\;\;
\co{17}=\frac{5 \lambda }{36 (\lambda
   -1)},\nonumber\\
&&\co{18}=\frac{1}{144} (-4 \lambda -4
   (\lambda +1)+16),\nonumber\\
&&\co{19}=\frac{16 (\lambda -1)-12 \lambda
   +4}{144 (\lambda -1)}.
\end{eqnarray}

\bibliography{PeculiaritiesBibPRDrev}

\end{document}